\documentclass[a4paper,dvips 2t,superscriptaddress,showkeys]{revtex4-1} 
\usepackage{epsfig,graphics,graphicx,amsmath,amssymb,float}

\usepackage{easyReview}

\begin{document}
	
	\title{\bf Effects of quantum geometric phase of a particle in an oscillating hard-wall spherical trap }
	\author{Reza Moazzemi }\email{r.moazzemi@qom.ac.ir}\affiliation{Department of Physics, University of Qom, Ghadir Blvd., Qom 371614-6611, I.R. Iran}\author{Seyed Mahdi Fazeli }\affiliation{Department of Physics, University of Qom, Ghadir Blvd., Qom 371614-6611, I.R. Iran}

	\begin{abstract}
	
	We obtain the { geometric phase for states of a particle  in} a spherical infinite potential well with moving walls in two different cases; First, when the radius of the well increases (or decreases)  monotonically. Second, when the radius changes oscillatory. In the latter case, we have solved the Schr\"odinger equation and found its solutions approximately.  {We obtain the transition rate for the possible real situation in an acousto-optic case which can reveal the effect of the geometric phase.  We show that the absorption or radiation peaks will appear if the energy gap between the incident photon and the modified energy difference of two levels by the geometric phase, is equal to the integer multiple of oscillation frequency.} 		
	\end{abstract}

	\maketitle
\section{Introduction}
The Berry phase is a gauge invariant and geometrical phase which is firstly defined by Berry in 1984 \cite{Berry1984}.
As a matter of fact, many physical situations are contained of objects whose behavior is specified up
to a phase by certain parameters. Berry phase appears when the transversal of a closed path by the parameters, which at the end of the path they have returned to their initial values, leads to a change of phase whose magnitude depends only on the path (For a collection of papers on the subject, see \cite{wilczek}).

As a simple example consider a magnet precessing about a magnetic field. Here the parameters are the three components of the field. The field is slowly rotated and finally, returns to its original direction. In this situation, the phase angle of the precession will be shifted 
from what it would be if the field had been held constant \cite{Cina1986}. In addition, recently many works have been published on the Berry phase and its applications in various fields. For example, in Ref. \cite{marija}  a system of strongly interacting one-dimensional (1D) bosons on a ring pierced by a synthetic magnetic flux tube is studied. In Ref. \cite{Antti}, Employing the spectral interference law of electromagnetic waves, a general expression for the Berry phase in Young's two-pinhole setup is established.

The most of
applications of the Berry phase are in solid state physics, especially in macroscopic phenomena which are
slow in time and smooth in space in comparison with the atomic scales (for review see \cite{Xiao2010}). For example, due to the Berry phase, when an electron completes a cycle around the Dirac point (a particular location in graphene's electronic structure), the phase of its wave function changes by $\pi$. Scanning tunneling spectroscopy revealed sudden jumps in conductivity as the external magnetic field was increased past a threshold value \cite{science}. As another important application of optical manipulation of the Berry phase in a solid-state spin qubit, in Ref. \cite{nature} authors demonstrate an all-optical method to accumulate a geometric phase, in an individual nitrogen–vacancy centre in diamond. Also, the measuring  Berry phase of graphene from wavefront dislocations in Friedel oscillations has been done recently in Ref. \cite{nature2}. In Quantum field theory  the subject of the Berry phase is also considered \cite{baggio,hsin}.

The solutions of time-dependent Schr\"{o}dinger equation for an infinite square-well potential with a moving wall in one dimension are known \cite{Doescher1969}. Here we derive its solutions in three dimensions for a spherical infinite potential well with moving wall, which also has been reported previously in Ref. \cite{musavi}. Then we calculate the {geometric} phase for this case.

The other three-dimensional case which we consider in this paper is the oscillatory moving wall, i.e.  we assume the radius of potential well changes oscillatory. In both cases, the changes in the radius of the well should occur slow enough so that we can use the adiabatic theorem. In the latter case, the solutions of Schr\"{o}dinger equation are found approximately. We then calculate the geometrical phase for this situation. Finally, we consider a more realistic situation in which a system oscillates with an acoustic  wave. For this case, we derive the {transition rate} and discuss the absorption and radiation modes. Note that, in this problem, we have only one time-dependent parameter, the radius of the well. However, this does not mean that the geometric phase is irrelevant in this problem. In fact, here, we present a way to measure some manifestations of geometric phase for this problem.


\section{Moving with constant velocity}\label{sec2}
\subsection{Solutions of Schr\"odinger Equation}

In this section we first find the solutions of the time dependent Schr\"{o}dinger equation for an infinite spherical well in which the radius of the well move with the constant velocity $v$. The time-dependent Schr\"{o}dinger equation is
\begin{equation}\label{sch}
	H(\mathbf{r},t){\Phi _{nlm}}(\mathbf{r},t) = i\hbar \frac{{\partial {\Phi _{nlm}}(\mathbf{r},t)}}{{\partial t}},
\end{equation} 
where $H(\mathbf{r},t)=-\frac{\hbar^2}{2m}\nabla^2+V(\mathbf{r},t)$ is the time-dependent Hamiltonian with
\begin{equation}\label{pot}
	V(\mathbf{r},t) = \left\{ \begin{array}{l}
		0{\qquad\ \, }r < a(t){\rm{ }}\\
		\infty {\qquad}r \ge a(t).{\rm{   }}
	\end{array} \right.
\end{equation}
{where $a(t)=a_0+vt$}. The general solution of the problem has the following form:
\begin{equation}
	\Psi(\mathbf{r},t)=\sum_{nlm}c_{nlm}\Phi _{nlm}(\mathbf{r},t).
\end{equation}
Since the potential is spherically symmetric we can easily separate the angular part of the solutions $\Phi _{nlm}(\mathbf{r},t)$, and write it down as
\begin{equation}
	\Phi _{nlm}(\mathbf{r},t) =\frac{u_{nl}(r,t)}rY_l^m(\theta ,\varphi )\,{e^{i\phi(r,t)}}
\end{equation}
where ${{\phi(r,t)}}$ is some phase, $Y_l^m(\theta ,\varphi )$ are the spherical harmonics and $u_{nl}(r,t)$ satisfy the Schr\"{o}dinger radial equation
\begin{equation}
	- \frac{{{\hbar ^2}}}{{2m}}\frac{{{\partial ^2}{u_{nl}}(r,t)}}{{\partial {r^2}}} + \frac{{{\hbar ^2}}}{{2m}}\frac{{l(l + 1)}}{{{r^2}}}{u_{nl}}(r,t) = i\hbar \frac{{\partial {u_{nl}}(r,t)}}{{\partial t}}.
\end{equation} 
On the other hand, using\textit{ adiabatic theorem} we know that the solutions should have the following form
\begin{eqnarray}\label{sol}
	{\Phi _{nlm}}(\mathbf{r},t) = \frac{1}{{j_{l + 1}({\beta _{nl}})}}\sqrt {\frac{2}{{{a(t)^3}}}} {j_l}[{\beta _{nl}}r/a(t)]\nonumber\\\times Y_l^m(\theta ,\varphi )\,{e^{i{\theta _{nl}}(t)}}{e^{if(r,t)}},
\end{eqnarray}
{with the energies ${E_{nl}}(t) = {{{\hbar ^2\beta _{nl}^2}}}/{{2m{a(t)^2}}}$ (similar to a well with radius of constant $a$). { The requrement of the appling adiabatic theorem is that the speed of the walls is much less than the characteristic speed of the particle i.e. $v\ll \frac{\hbar}{ma}$.} In \eqref{sol}, $j_l(x)$ is the spherical Bessel function of order $l$, $\beta_{nl}$ is $n$th zero of the $l$th spherical Bessel function. Now the dynamical phase is defined by}
\begin{eqnarray}
	{\theta _{nl}}(t)& = & - \frac{1}{\hbar }\int_0^t {{E_{nl}}(t')} dt'\label{dynp}\\
	& = &  - \frac{{\hbar \beta _{nl}^2}}{{2mv}}\left( {\frac{1}{{{a_{\,0}}}} - \frac{1}{{{a_{\,0}} + vt}}} \right).
\end{eqnarray}
To find $f(r,t)$ in \eqref{sol}, we try to solve the Eq. \eqref{sch} by this solution. If we solve the real and imaginary part separately, we find that
\begin{equation}
	f\left( {r,t} \right) = \frac{{mv{r^2}}}{{2\hbar \left( {{a_0} + vt} \right)}}.
\end{equation}
Therefore the solution of the Schrodinger equation for infinite spherical well with moving wall is
\begin{eqnarray}
	{\Phi _{nlm}}(\mathbf{r},t) &=&
	{\varphi _{nlm}}(\mathbf{r},t){e^{i{\theta _{nl}}(t)}}\nonumber\\
	&=&
	\frac{1}{{j_{l + 1}({\beta _{nl}})}}{\sqrt {\frac{2}{{{{({a_{0}} + vt)}^3}}}} }\,\,{j_l}\left[ {\frac{{{\beta _{nl}}r}}{{{a_{0}} + vt}}} \right]\nonumber\\&&\hspace{-2.3cm}\times{\exp{\left[i {\frac{{mv{r^2}}}{{2\hbar \left( {{a_{0}} + vt} \right)}} -i \frac{{\hbar \beta _{nl}^2}}{{2mv}}\left( {\frac{1}{{{a_{0}}}} - \frac{1}{{{a_{0}} + vt}}} \right)} \right]}}Y_l^m(\theta ,\varphi ).\nonumber\\
\end{eqnarray}
These solutions are in agreement with the ones reported in Ref. \cite{musavi}.
\subsection{The {geometric} phase}

Now, we are ready to obtain the geometric phase 
\begin{equation}\label{berryp}
	{\gamma _{nl}}(t) = i\int_0^t {\left\langle {{\varphi _{nlm}}(\mathbf{r},t')\left| {\frac{\partial }{{\partial t'}}{\varphi _{nlm}}(\mathbf{r},t'} \right.)} \right\rangle \,} dt'.
\end{equation}
{The Berry phase is the above geometric phase when the system returns to its original state (or to one of its degenerate state) in a cyclic path 
\begin{equation}\label{berryph.}
	{\gamma _{nl}}(C) = i\oint_C {\left\langle {{\varphi _{nlm}}(\mathbf{r},t')\left| {\frac{\partial }{{\partial t'}}{\varphi _{nlm}}(\mathbf{r},t'} \right.)} \right\rangle \,} dt'.
\end{equation}}
Computing the inner product of \eqref{berryp} we have
\begin{eqnarray}
	\left\langle {{\varphi _{nlm}}\left| {\frac{{\partial {\varphi _{nlm}}}}{{\partial t'}}} \right.} \right\rangle  = \int_0^a {\varphi _{nlm}^*(r,t)} \frac{\partial }{{\partial t'}}\varphi _{nlm}^{}(r,t){r^2}drd\Omega\nonumber\\
	 =  - i\frac{{mv^2}}{{6\hbar \beta _{nl}^2}}{\left[ {\frac{{j_{l - 1}^{}\left( {{\beta _{nl}}} \right)}}{{j_{l + 1}^{}\left( {{\beta _{nl}}} \right)}}} \right]^2}\left[ {4l(l + 1) - 3 + 2\beta _{nl}^2} \right].
\end{eqnarray}
Therefore,
\begin{eqnarray}\label{gp..}
	{\gamma _{nl}}(t) = \frac{{mv}}{{6\hbar \beta _{nl}^2}}{\left( {\frac{{j_{l - 1}^{}\left( {{\beta _{nl}}} \right)}}{{j_{l + 1}^{}\left( {{\beta _{nl}}} \right)}}} \right)^2}\left[ {4l(l + 1) - 3 + 2\beta _{nl}^2} \right]\nonumber\\\times\left[ {a(t) - {a_{\,0}}} \right].
\end{eqnarray}
{The appearance of $v$ in the above relation shows how this geometric phase depends on the path taken.} We can write \eqref{gp..}  as
\begin{eqnarray}
	{\gamma _{nl}}(\mathbf{a}) = f_{nl}a+\gamma_0,
\end{eqnarray}
where $f_{nl}$ and $\gamma_0$ are constant.
{It is clear that the above time dependent geometric phase directly depends on the radius of the well which is a geometric parameter.
}

\section{The oscillatory moving wall}
\subsection{Solutions of the Schr\"odinger equation }
In this section we derive the Berry phase when the wall of the well is moving oscillatory. Again the potential is the same as Eq. \eqref{pot} but with $a(t)=a_0+b\sin(\omega t)$. The dynamical phase is now 
\begin{eqnarray}
\nonumber\hspace{0cm}	\theta_{nl}^{\rm{osc}}(t)&=&-\frac{1}{\hbar}\int^t \frac{\hbar^2 \beta_{nl}^2}{2 m (a_0+b\sin \omega t')^2} \, dt'
\\&	=&-\frac{\hbar\beta_{nl}^2}{2m\omega}\Bigg\{\frac{2a_0\arctan\left[\frac{b+a_0\tan(\omega t/2)}{\sqrt{a_0^2-b^2}}\right]}{(a_0^2-b^2)^{3/2}}\nonumber\\&&+\frac{b\cos\omega t}{(a_0^2-b^2)(a_0+b\sin \omega t)}\Bigg\}+\phi_0
\\&	=&{-\frac{\bar{E}_{nl}}{\hbar}t+\zeta_{nl}(t)+\phi_0},
\end{eqnarray}
where {$\bar{E}_{nl}$ is the averaged energy in one period, $\zeta_{nl}(t)$ some periodic function of $t$} and  $\phi_0$ a constant phase. We try the following solution: 
\begin{eqnarray}\label{sol2}
	{\Phi _{nlm}^{\rm{osc}}}(\mathbf{r},t) = \frac{1}{{j_{l + 1}({\beta _{nl}})}}\sqrt {\frac{2}{{{a(t)^3}}}} {j_l}[{\beta _{nl}}r/a(t)]Y_l^m(\theta ,\varphi )\nonumber\\\times e^{i{\theta_{nl}^{\rm{osc}}(t)}}{e^{ig(r,t)}},
\end{eqnarray}
in the Schr\"odinger equation \eqref{sch}. We encounter two separate equations for imaginary and real parts. To satisfy the imaginary part it is sufficient that these two following relations hold:
\begin{eqnarray}\label{difeqs}
	\left\{\begin{array}{l}
		\hbar(a_0+b\sin\omega t)\dfrac{\partial g(r,t)}{\partial r}-bm\omega r\cos\omega t=0\vspace{.2cm}\\
		\hbar(a_0+b\sin\omega t)\left(r\dfrac{\partial^2 g(r,t)}{\partial r^2}+2(l+1)\dfrac{\partial g(r,t)}{\partial r}\right)\\\hspace{2.5cm}-bm\omega r(2l+3)\cos\omega t=0
	\end{array}\right.
\end{eqnarray}
The solution of the first differential equation in \eqref{difeqs} is
\begin{equation}\label{gxt}
	g(r,t)=\frac{bm\omega r^2\cos\omega t}{2\hbar(a_0+b\sin\omega t)}+h(t),
\end{equation}
where $h(t)$ is some function of time. By the way, this result satisfies also the second equation. But now, for the real part it is necessary that following differential equation holds:
\begin{eqnarray}\label{realp}
	\frac{\hbar}{2m}\left(\dfrac{\partial g(r,t)}{\partial r}\right)^2+ \dfrac{\partial g(r,t)}{\partial t}=0
\end{eqnarray} 
This equation leads us to take $h(t)=0$. Unfortunately, the resultant $g$ does not satisfy the Eq. \eqref{realp} exactly. However, it can be approximately hold if
\begin{equation}
\frac{mb\omega^2r^2\sin\omega t}{2a{(t)}}\approx0.
\end{equation}
When we compare the above expression with a particle energy level, $\hbar^2\beta_{nl}^2/2ma^2$, we find
\begin{equation}\label{req}
	\omega\ll {\frac{\hbar\beta_{nl}}{ma_0^2}\sqrt{\frac{a_0}{b}}\sim 10^{-2}\mbox{[m$^2$/S]}a_0^{-2}}
\end{equation}
{Note that this condition is weaker than adiabatic theorem requirement $b\omega\ll \frac{\hbar}{ma}$.} Satisfying the requirement \eqref{req}, the solution of the Scher\"odinger equation for an infinite oscillatory spherical well can be written as
\begin{eqnarray}
&&	{\Phi _{nlm}^{\rm{osc}}}(\mathbf{r},t) =
	{\varphi _{nlm}^{\rm{osc}}}(\mathbf{r},t){e^{i{\theta _{nl}^{\rm{osc}}}(t)}}\nonumber\\
	&&\approx
	\frac{1}{{j_{l + 1}^{}({\beta _{nl}})}}{\left( {\frac{2}{{{{({a_{\,0}} + b\sin\omega t)}^3}}}} \right)^{1/2}}{j_l}\left( {\frac{{{\beta _{nl}}r}}{{{a_{\,0}} + b \sin\omega t}}} \right)\nonumber\\&&\times{\exp\left[i\frac{bm\omega r^2\cos\omega t}{2\hbar(a_0+b\sin\omega t)}+i\theta_{nl}^{\rm{osc}}(t) \right]}\,Y_l^m(\theta ,\varphi ).
\end{eqnarray}

\subsection{The {geometric} phase}
To obtain the geometric phase, again we use the Eq. \eqref{berryp}. In this case we need to the following inner product
\begin{eqnarray}
	\left\langle {{\varphi _{nlm}^{\rm{osc}}}\left| {\frac{{\partial {\varphi _{nlm}^{\rm{osc}}}}}{{\partial t'}}} \right.} \right\rangle=-i\frac{  b m \omega^2}{{12 \hbar \beta_{nl}^2} }\left(4 l^2+4 l-3+2 \beta_{nl}^2\right) \nonumber\\\times j_{l-1}^2\left(\beta_{nl}\right) (a_0+b \sin  \omega t).
\end{eqnarray}
Note that the integral of the above expression is not elementary. We use the following integral relation
\begin{eqnarray}
&&	\int {{x^4}{j_l}{{\left( x \right)}^2}dx }=
\\&& \frac{{\pi {x^{2l + 5}}{_2}{F_3}\left( {l + 1,l + \frac{5}{2};l + \frac{3}{2},l + \frac{7}{2},2l + 2; - {x^2}} \right)}}{{{2^{2l + 2}}\left( {2l + 5} \right)\Gamma {{\left( {l + \frac{3}{2}} \right)}^2}}}  + C\nonumber\\
	&&= \frac{1}{{12}}\left( { - 2l - 3} \right){x^2}\left( {4{l^2} + 2{x^2} - 1} \right){j_{l - 1}}\left( x \right){j_l}\left( x \right)\nonumber\\
&&\hspace{1cm}	+ \frac{1}{{12}}{x^3}\left( {4{l^2} + 8l + 2{x^2} + 3} \right){j_l}{\left( x \right)^2}\nonumber\\&& \hspace{1cm}+ \frac{1}{{12}}{x^3}\left( {4{l^2} + 4l + 2{x^2} - 3} \right){j_{l - 1}}{\left( x \right)^2} + C,
\end{eqnarray}
where ${_2}{F_3}(a;b;x)$ is the generalized hypergeometric function and $\Gamma (x)$ the gamma function. Then, for the {geometric} phase we have
\begin{eqnarray}\label{geo.}
	\gamma _{nl}^{\rm{osc}}(t)&=&\frac{  mb \omega}{{12 \hbar \beta_{nl}^2} }\left(4 l^2+4 l-3+2 \beta_{nl}^2\right) \nonumber\\&&\qquad{\times j_{l-1}^2}\left(\beta_{nl}\right) \left[b\omega t+a_0 (1-\cos  \omega t)\right]
\\ &=&{-\frac{\varepsilon_{nl}}{\hbar}t+\zeta'_{nl}(t)},\label{geom.}
\end{eqnarray}
{where \[\varepsilon_{nl}= \frac{-mb^2 \omega^2}{{12 \beta_{nl}^2}} j_{l-1}^2\left(\beta_{nl}\right)\left(4 l^2+4 l-3+2 \beta_{nl}^2\right)\] 
 \[\zeta'_{nl}(t)=\frac{mb \omega a_0}{{12\hbar   \beta_{nl}^2}} j_{l-1}^2\left(\beta_{nl}\right)\left(4 l^2+4 l-3+2 \beta_{nl}^2\right)(1-\cos\omega t).\]
{The second term in Eq. \eqref{geom.} is some periodic function and therefore the first term is just the Berry phase if we put $t=T=\frac{2\pi}{\omega}$.} In Fig. \ref{fig1} we have plotted the dynamical, geometrical and total phases separately for some first quantum numbers. We have illustrated the ratio of geometric and dynamic phases in Fig. \ref{fig2}.

\begin{figure}[H]
\centering	\includegraphics[width=8.5cm]{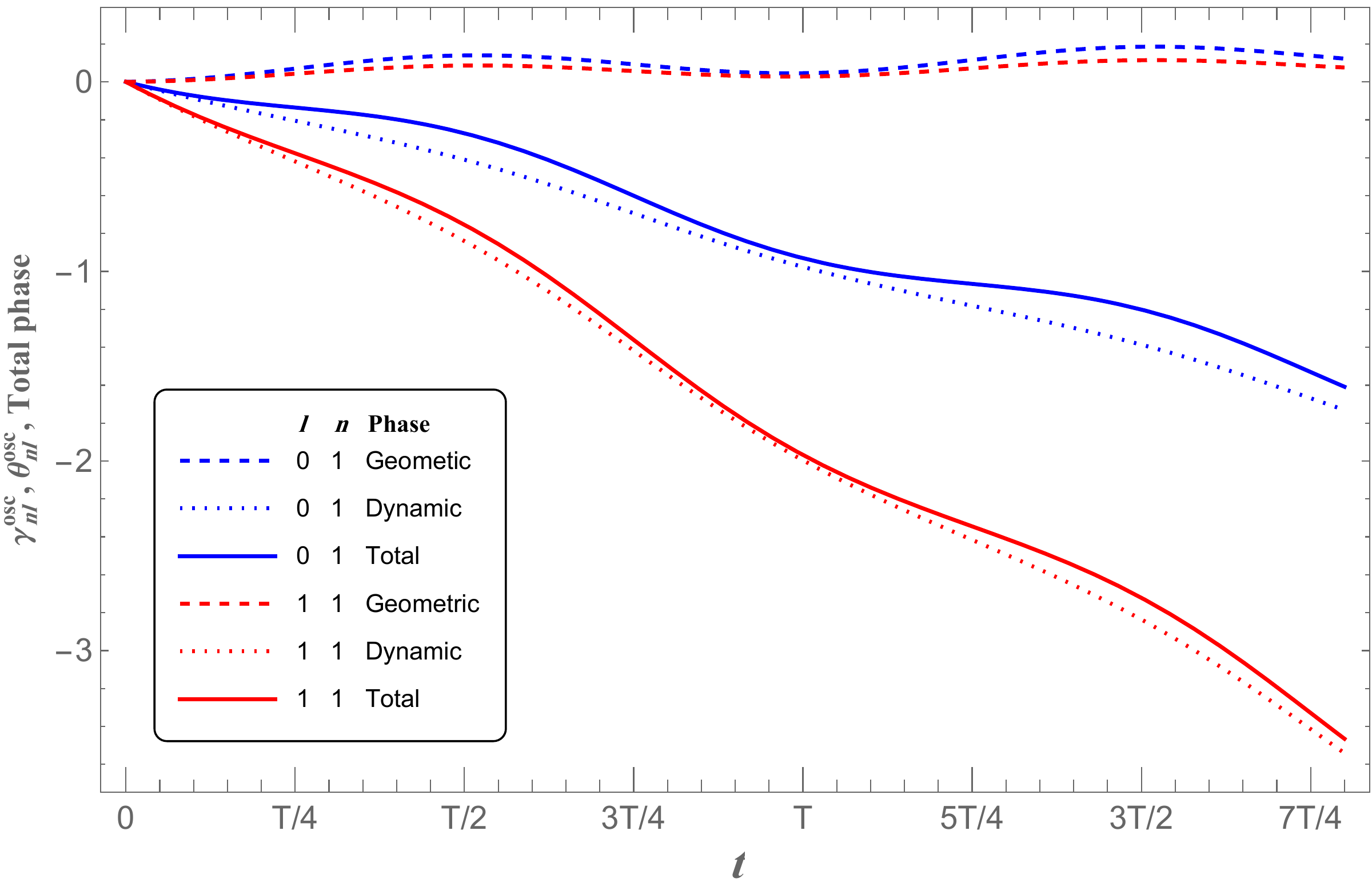} \caption{ {\small  The dynamical, geometrical and total phases.}}
	\label{fig1}
\end{figure}

\begin{figure}[H]
\centering	\includegraphics[width=8.5cm]{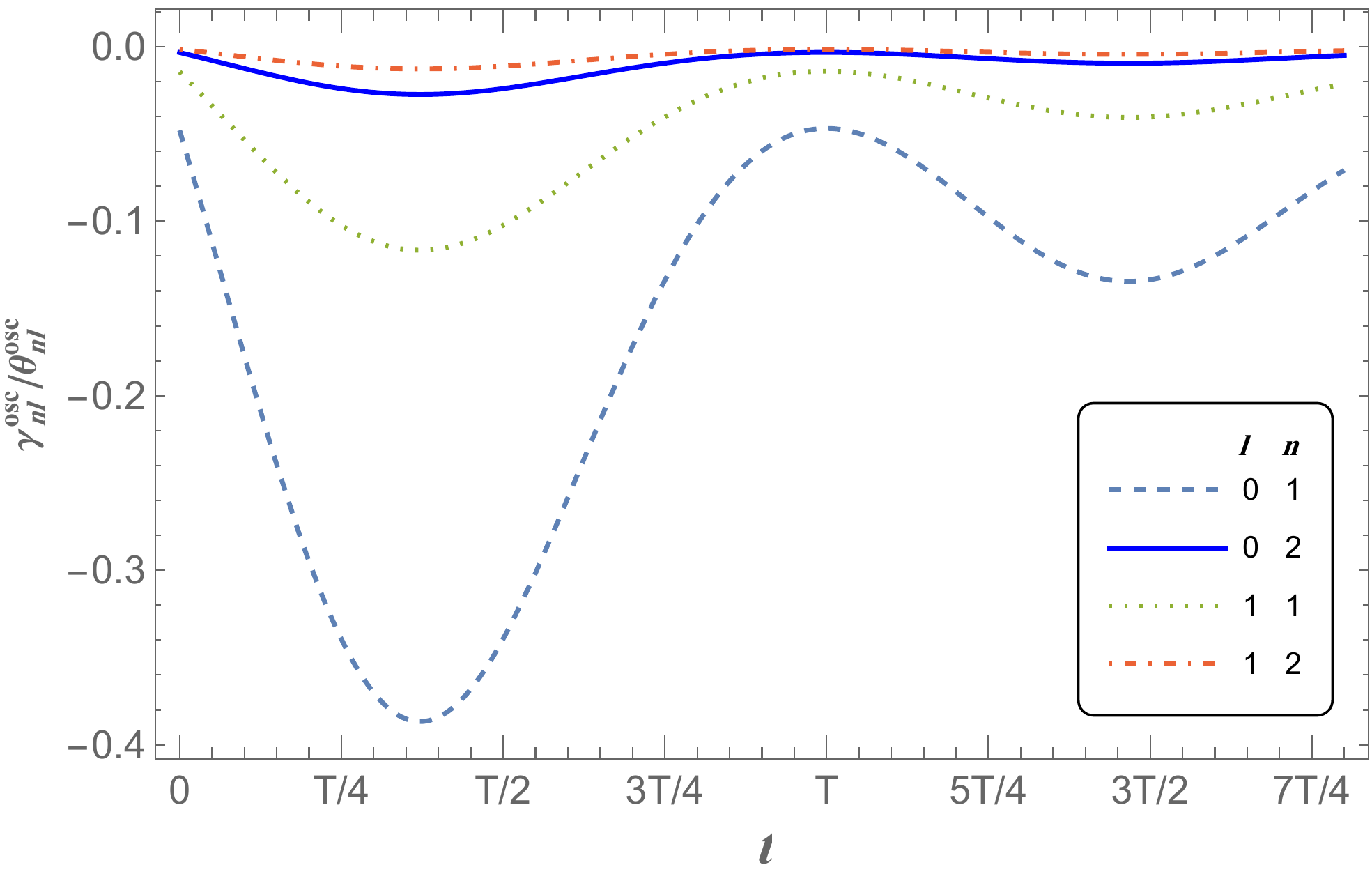} \caption{ {\small The ratio of geometric and dynamic phases.}}
	\label{fig2}
\end{figure}

\section{Acousto-optic example}

Consider a metal sphere that fluctuates sinusoidally under an elastic acoustic wave. {Such waves can be easily created with frequencies much more than MHz, so that for a well radius about 500 nm, the condition of Eq.\eqref{req} is satisfied.} For the approximation of the infinite  potential well to be acceptable, the metal must have a high work function.  This structure can be considered as an example of a soluble system. Now, we have
\begin{equation}
	H_0(t)|\phi_{nlm}^{\rm{osc}}\rangle=i\hbar\frac{\partial}{\partial t}|\phi_{nlm}^{\rm{osc}}\rangle.
\end{equation}
Here,
\begin{equation}
	|\phi_{nlm}^{\rm{osc}}\rangle=|\phi^0_{nlm}\rangle e^{-i\eta_{nl}},
\end{equation}
where { $\eta_{nl}$ is the total phase (dynamical plus geometrical, see Eq. \eqref{geom.})
	\begin{eqnarray}
		\eta_{nl}&=&-\frac{\bar{E}_{nl}}{\hbar}t-\frac{\varepsilon_{nl}}{\hbar}t+\zeta_{nl}+\zeta'_{nl}\label{outetild}
	\end{eqnarray} 
	  and $|\phi^0_{nlm}\rangle$ is the rest of wave function. 
	{  We can \textit{generally} write Eq. \eqref{outetild}  in the form 
	  	\begin{eqnarray}
		\eta_{nl}=-\frac{\tilde{E}_{nl}}{\hbar}t+\tilde{\zeta}_{nl},\label{etild}
	\end{eqnarray} 
	  where, $\tilde{E}_{nl}=\bar{E}_{nl}+\varepsilon_{nl}$ that we call modified time-averaged energy in one period and $\tilde{\zeta}_{nl}=\zeta_{nl}+\zeta'_{nl}$ is some periodic function.} Now, if we expose this system to electromagnetic fields, we can consider the Hamiltonian of the system as $H(t) = H_0 + \lambda V(t)$. Considering
\begin{equation}
	|\psi\rangle=\sum_{nlm}c_{nlm}(t)|\phi_{nlm}^{\rm{osc}}\rangle
\end{equation}
we can solve the Schr\"odinger equation to have
\begin{equation}
	i\hbar\frac{\partial}{\partial t}c_{nlm}(t)=\sum_{n'm'l'}c_{n'l'm'}(t)\langle\phi_{nlm}^{\rm{osc}}|\lambda V(t)|\phi_{n'l'm'}^{\rm{osc}}\rangle.
\end{equation}
{Note that, one can easily show, that for a typical metal the penetration depth of the electric field is much more than the dimensions of the metal sphere.} On the other hand, since $\lambda V = - e \bf E \cdot r$, we simply take the field in the $z$ direction, so that $\lambda V = - e E r \cos \theta$. Therefore,
\begin{equation}
	\langle\phi_{nlm}|\lambda V|\phi_{n'l'm'}\rangle=\frac{a_0+b\sin\omega t}{a_0}\langle\phi_{nlm}^{0}|\lambda V(t)|\phi_{n'l'm'}^{0}\rangle,
\end{equation}
where $|\phi_{n'l'm'}^{0}\rangle$ is time independent for constant sphere ($b=0$). Consequently we get
\begin{eqnarray}
&&\hspace{-.9cm}	i\hbar\frac{\partial}{\partial t}c_{nlm}(t)=\sum_{n'm'l'}c_{n'l'm'}(t)\frac{a_0+b\sin\omega t}{a_0}e^{-i(\eta_{n'l'}-\eta_{nl})}\nonumber\\&&\hspace{3cm}\times\langle\phi_{nlm}^{0}|\lambda V(t)|\phi_{n'l'm'}^{0}\rangle.
\end{eqnarray}
Now, if at initial point, the electron is in $n_0l_0m_0$ level, then $c_{nlm}(0)=\delta_{nn_0}\delta_{ll_0}\delta_{mm_0}$ and up to ${\cal O}(\lambda)$ we have
\begin{eqnarray}
&&\hspace{-.9cm}	i\hbar\frac{\partial}{\partial t}c_{nlm}(t)=\frac{a_0+b\sin\omega t}{a_0}e^{-i(\eta_{n_0l_0}-\eta_{nl})}\nonumber\\&&\hspace{3cm}\times\langle\phi_{nlm}^{0}|\lambda V(t)|\phi_{n_0l_0m_0}^{0}\rangle.
\end{eqnarray}
and accordingly,
\begin{eqnarray}
&&\hspace{-1.4cm}	c_{nlm}(t)=\frac{1}{i\hbar}\int_0^t\frac{a_0+b\sin\omega t'}{a_0}e^{-i\Delta\eta_{nl}}
\nonumber\\&&\hspace{2.5cm}\times
\langle\phi_{nlm}^{0}|\lambda V(t')|\phi_{n_0l_0m_0}^{0}\rangle dt',
\end{eqnarray}
{where $\Delta\eta_{nl}=\eta_{n_0l_0}-\eta_{nl}$. Since the electric field is periodic and Hermitian operator, we can write $V(t)=V_0e^{-i\omega_{\rm{ph}}t}+V_0^\dagger e^{i\omega_{\rm{ph}}t}$, where $\omega_{\rm{ph}}$ is the frequency of photon. On the other hand, since  $\tilde{\zeta}_{nl}(t)$ is a periodic function with angular frequency $\omega$,} we can write the oscillating terms in Fourier series form as follows:
\begin{equation}
	\frac{a_0+b\sin\omega t}{a_0}e^{-i\tilde{\zeta_{nl}}(t)}=\sum_{k=-\infty}^{\infty} f^k_{nl}e^{-ik\omega t}.
\end{equation}
Therefore,
\begin{eqnarray}
	c_{nlm}(t)&=&\frac{1}{i\hbar}\sum_kf^k_{nl}\bigg[\langle\phi_{nlm}^{0}|\lambda V_0|\phi_{n_0l_0m_0}^{0}\rangle\int_0^te^{-i(\Delta\tilde{E}_{nl}+\hbar\omega_{\rm{ph}}+\hbar k\omega)t/\hbar } dt'\nonumber\\&&\hspace{1.6cm}+\langle\phi_{nlm}^{0}|\lambda V_0^\dagger|\phi_{n_0l_0m_0}^{0}\rangle\int_0^te^{-i(\Delta\tilde{E}_{nl}-\hbar\omega_{\rm{ph}}+\hbar k\omega)t/\hbar } dt'\bigg]\nonumber\\
	&=&\frac{1}{i\hbar}\sum_kf^k_{nl}\bigg\{\langle\phi_{nlm}^{0}|\lambda V_0|\phi_{n_0l_0m_0}^{0}\rangle e^{-i(\Delta\tilde{E}_{nl}+\hbar\omega_{\rm{ph}}+\hbar k\omega)t/2\hbar }\mbox{sinc}\left[(\Delta\tilde{E}_{nl}+\hbar\omega_{\rm{ph}}+\hbar k\omega)t/2\hbar\right]t\nonumber\\&&\hspace{.5cm}
	+\langle\phi_{nlm}^{0}|\lambda V^\dagger_0|\phi_{n_0l_0m_0}^{0}\rangle e^{-i(\Delta\tilde{E}_{nl}-\hbar\omega_{\rm{ph}}+\hbar k\omega)t/2\hbar }\mbox{sinc}\left[(\Delta\tilde{E}_{nl}-\hbar\omega_{\rm{ph}}+\hbar k\omega)t/2\hbar\right]t
	\bigg\}.\label{cnlm}
\end{eqnarray} \vspace{-.8cm}
To have the probability, we should multiply Eq. (\ref{cnlm}) in its complex conjugate. At large $t$ we have \vspace{.5cm}
\[\left[t\ \mbox{sinc}(tx/2)\right]^2=\frac{\sin^2(tx/2)}{(x/2)^2}\cong2\pi\delta(x)t.
\]
 For the transition rate we have
\begin{eqnarray}
	\Gamma_{n_0l_0m_0\to nlm}&=&\frac{2\pi}{\hbar^2}\sum_k|f^k_{nl}|^2\Bigg\{|\langle\phi_{nlm}^{0}|\lambda V_0|\phi_{n_0l_0m_0}^{0}\rangle|^2\delta\left[\frac{\Delta\tilde{E}_{nl}}{\hbar}+\omega_{ph}+ k\omega\right]\nonumber\\&&\hspace{2cm}+|\langle\phi_{nlm}^{0}|\lambda V_0^\dagger|\phi_{n_0l_0m_0}^{0}\rangle|^2\delta\left[\frac{\Delta\tilde{E}_{nl}}{\hbar}-\omega_{ph}+ k\omega\right]\Bigg\}.\label{tr}
\end{eqnarray}
This is just the {transition rate} for dielectric radiation modified with respect to oscillation. From Eq. \eqref{tr} we can 
understand that if the energy gap between the incident photon $\omega_{\rm{ph}}$, and difference of the {modified} energy of two levels $\Delta\tilde{E}_{nl}$, is equal to the integer multiple of oscillation frequency $\omega$, we will also have absorption or radiation peaks. {Here, the effect of geometric phase is hidden in the $\Delta\tilde{E}_{nl}$.}

\section{Summary}
The geometric phase for spherical infinite potential well with moving wall in two different cases has been considered. First, when the radius of well increases (or decreases)  monotonically. The counterpart of this problem in one dimension is standard and well-known. We have shown that there is a way to observe some manifestations of geometric phase, though the parameter space of this problem is one-dimensional. To solve this problem we have, at first, found the wave functions which satisfy the Schr\"odinger equation. Second, when the radius changes oscillatory. This problem is closer to physical situations. For this case, we have also solved the Schr\"odinger equation and found its solutions approximately. For both cases, we compute the geometrical phases. In the other section, {we have derived the transition rate for a possible real situation which manifests the effect of geometric phase} .  We have shown that, in an acousto-optic case, the absorption or radiation peaks will appear if the energy gap between the incident photon and the {difference of modified averaged energy  of two levels due to Berry phase}, is equal to the integer multiple of oscillation frequency.




\vspace{-.5cm}
\begin{thebibliography}{0}



\bibitem{Berry1984}
M V Berry, \textit{P. Roy. Soc A-Math. Phy.} \textbf{392}, 45 (1984) 


\bibitem{wilczek}
 A Shapere and F Wilczek,
\textit{Geometric Phases in Physics} (World Scientific, Singapore, 1989)

\bibitem{Cina1986}
 J A Cina, \textit{Chem. Phys. Lett.} \textbf{132}, {393}  (1986)

\bibitem{Xiao2010}
D Xiao, M C Chang  and Q Niu,\textit{ Rev. Mod. Phys.} \textbf{82}, {1959}  ({2010})

\bibitem{marija}
M Todori\'c, B Klajn, D Juki\'c and H Buljan
\textit{Phys. Rev. A} \textbf{102},  013322 (2020)


\bibitem{Antti}
A Hannonen, H Partanen, J Tervo, T Set\"al\"a and A.T. Friberg
\textit{Phys. Rev. A} \textbf{99},  053826 (2019)

\bibitem{science}
F Ghahari, et al. 
\textit{Science} \textbf{356},  845 (2017)

\bibitem{nature}
C G Yale, F J Heremans, B B Zhou, A Auer, G Burkard and D D  Awschalom,
\textit{ Nat. photonics} \textbf{10}, 184  (2016)

\bibitem{nature2}
C Dutreix, H González-Herrero, I Brihuega, M I Katsnelson, C Chapelier and V T Renard,
\textit{Nature} \textbf{574},  219 (2019)

\bibitem{baggio}
M Baggio,  V Niarchos and K Papadodimas. 
\textit{J. high Energy Phys.} \textit{04}, 1 (2017) 

\bibitem{hsin}
P S  Hsin, A Kapustin and R Thorngren, 
 \textit{Phys. Rev. B} \textbf{102},  245113 (2020).

\bibitem{Doescher1969}
S W  Doescher and M H Rice, Am. J. Phys. \textbf{37},  {1246} ({1969})


\bibitem{musavi}
S V Mousavi, 
\textit{Europhys. Lett. }\textbf{99}, 30002  (2012)

\end{thebibliography}
\end{document}